# FROM LOCAL VELOCITIES TO MICROWAVE BACKGROUND

Saleem Zaroubi[1], Naoshi Sugiyama[2], Joseph Silk[1],
Yehuda Hoffman[3], & Avishai Dekel[1,3]

[1] Department of Astronomy and Center for Particle Astrophysics,
University of California, Berkeley, CA 94720
[2] Department of Physics, Kyoto University, Kyoto 606, Japan
[3] Racah Institute of Physics, The Hebrew University, Jerusalem 91904, Israel

## ABSTRACT

The mass density field as extracted from peculiar velocities in our cosmological neighborhood is mapped back in time to the CMB in two ways. First, the density power spectrum ($P_k$) is translated into a temperature angular power spectrum of sub-degree resolution ($C_l$) and compared to observations. Second, the local density field is translated into a temperature map in a patch on the last-scattering surface of a distant observer.

A likelihood analysis of the Mark III peculiar velocity data have constrained the range of parameters for $P_k$ within the family of COBE-normalized CDM models (Zaroubi *et al.* 1996), favoring a slight tilt in the initial spectrum, $n < 1$. The corresponding range of $C_l$'s is plotted against current observations, indicating that the CMB data can tighten the constraints further: only models with *small* tilt ($n \sim 0.9$) and *high* baryonic content ($\Omega_b \sim 0.1$) could survive the two data sets simultaneously.

The local mass density field that has been recovered from the velocities via a Wiener method is convolved with a Boltzmann calculation to recover $10'$-resolution temperature maps as viewed from different directions. The extent of the CMB patch and the amplitude of fluctuations depend on the choice of cosmological parameters, *e.g.*, , the local $100 h^{-1} Mpc$ sphere corresponds to $90'$ to $30'$ at the CMB for $\Omega$ between 1 and 0 respectively. The phases of the temperature map are correlated with those of the density field, contrary to the contribution of the Sachs-Wolfe effect alone. This correlation suggests the possibility of an inverse reconstruction of the underlying density field from CMB data with interesting theoretical implications.

*Subject headings:* cosmic microwave background — cosmology: theory — cosmology: observation — dark matter — large-scale structure of universe

# 1. INTRODUCTION

Sub-degree angular scales on the last scattering surface of the CMB correspond to comoving scales of order $10 - 100 h^{-1} Mpc$ in the present local universe (LSS). The ongoing efforts to measure CMB anisotropies on sub-degree scales (see a review by White, Scott & Silk 1994) and mass-density and galaxy-density fluctuations in our cosmological neighborhood (see a review by Dekel 1994; Strauss & Willick 1995) provide independent probes for the structure on similar comoving scales at very different cosmological epochs. First, the comparison between the two is a unique way of testing the hypothesis of fluctuation growth by gravitational instability (GI). Then, under the assumption that the local neighborhood is a fair sample, the combined data can be used to constrain the parameters of competing cosmological models. Finally, the comparison could address the question of whether the local neighborhood is indeed a fair sample, as opposed, for example, to a local void.

A comparison between the CMB and LSS fluctuation fields field was suggested by Juszkiewicz, Gorski & Silk (1987). Bertschinger, Gorski & Dekel (1990, BGD) used the density field as derived by the POTENT procedure from peculiar velocities in our local neighborhood (Dekel *et al.* 1990; Bertschinger *et al.* 1990) to derive using GI a corresponding CMB temperature map as viewed by a distant observer. The advantage of using peculiar-velocity data as opposed to galaxy density surveys is that the former is a direct dynamical measurement free of the unknown galaxy-density biasing. Based on their analysis of the Sachs-Wolfe effect and a crude approximation for the "Doppler" effect, BGD predicted $\Delta T/T \gtrsim 10^{-5}$ for a beam of $30'$ FWHM, to be confirmed later by COBE.

The peculiar velocity data set have increased by factor 5 in the Mark III catalog, improved methods have yielded a more extended and reliable density field, and it's power spectrum has been computed in several different ways. This calls for a more accurate translation back in time into the CMB which we attempt here in two ways.

A likelihood analysis of the Mark III peculiar velocity data have constrained the allowed range of parameters of the mass power spectrum ($P_k$) within the framework of the CDM family of models under COBE normalization (Zaroubi *et al.* 1996). Under the assumption that the local neighborhood is a fair sample, and that gravity is the engine driving the fluctuation growth, we translate this range into a range in the angular power spectrum of temperature fluctuations at the CMB, $C_l$, which is then compared to recent CMB observations. The purpose of the simultaneous fit to the two data sets is to tighten the constraints on the parameters of the CDM family of models such as the cosmological density parameters $\Omega_m$ and $\Omega_\Lambda$, the baryonic contribution $\Omega_b$, and the power index on large scales $n$.

We then translate the local mass density field into a map of $\Delta T/T$ in the spirit of BGD but with a much improved accuracy, now taking into account the important physical effects on sub-degree scales that were treated very crudely before. The local mass density field has been reconstructed from the Mark III velocities via a Wiener Filter method (Zaroubi, Hoffman & Dekel 1997), which is an alternative to the POTENT procedure. It is then convolved with the output of a numerical calculation using the Boltzmann-equation to yield temperature maps with smoothing 10' FWHM as viewed from different directions. These maps contain phase information on top of the power spectrum, independent of any assumption concerning the Gaussianity of the fluctuations. We present the method and examine the effect of the cosmological model via the parameters $\Omega_m$, $\Omega_\Lambda$ and $\Omega_b$. A comparison with observed maps of the CMB at similar resolution may shed light on the question of whether we live in a typical neighborhood or possibly in a local void (Turner, Cen & Ostriker 1992). We also study the correlation between the temperature and density maps, which may serve for an inverse reconstruction of density fluctuations from a whole-sky map of CMB fluctuations.

The outline of the paper is as follows: In §2 we pursue the power spectrum analysis; from the Mark III peculiar velocity data, through the derived mass power spectrum, the pre-recombination evolution and the translation to $C_l$'s, ending with plots of the corresponding allowed range of $C_l$'s against CMB observations. In §3 we describe the density reconstruction via Wiener analysis and the mapping back in time of the the local neighborhood into a CMB patch. Our conclusions are discussed in §4.

## 2. POWER SPECTRA: $P_k$ versus $C_l$

### 2.1. Power Spectrum from Peculiar Velocities

The Mark III catalog (Willick *et al.* 1995, 1996, 1997) is composed of several Tully Fisher (TF) and $D_n - \sigma$ data sets of radial peculiar velocities in inferred-distance space. The major effort in assembling this catalog was to impose common selection criteria and to apply self-consistent calibration of the distance indicators in the different data sets. The merged set consists of about 3500 galaxies, mostly spirals. The galaxies were gathered into groups and clusters in order to reduce the scatter in the distances and the resulting Malmquist bias. The final catalog consists of $\sim 1200$ objects with redshifts, inferred distances, and estimated errors.

The power spectrum of mass-density fluctuations has been determined from the Mark III catalog via likelihood analysis (Zaroubi *et al.* 1996). The analysis is based on the assumption that both the peculiar velocities and their errors are Gaussian random fields, and that the relation between peculiar velocity and density is provided by the linear approximation to GI. The power spectrum is assumed to have a specific functional form with a few free parameters.

Among the models tested by Zaroubi *et al.* as priors for $P_k$, they explored the general CDM family of models, in which the variable parameters are the cosmological densities $\Omega_m$ and $\Omega_\Lambda$ (between 0.1 and 1.0), the Hubble constant $h$ (between 0.4 and 1.0), the power index $n$ (between 0.5 and 1.0), and the ratio of tensor to scalar fluctuations [assumed to be either zero or $T/S = 7(1 - n)$]. The power spectra were normalized by the 4-year data of COBE, where a value of $\Omega_b h^2 = 0.024$ was adopted from Tytler *et al.* (1996). The acceptable range of parameters at the $\sim 90\%$ confidence level can be crudely described by $\Omega_m h_{50}^{1.2} n^\nu = 0.8 \pm 0.2$, where $\nu = 3.4$ and 2.0 for $T/S = 7(1 - n)$ and zero respectively. [A result in the same ball park was obtained by White *et al.* (1996) based on the power spectrum of galaxy density by Peacock and Dodds (1994) and the 2-year data of COBE.]

Figure 1 shows the the best-fit CDM power spectra with and without tensor fluctuations. It also displays the power spectrum as computed by Kolatt and Dekel (1997) from the POTENT smoothed density field recovered from the same Mark III data independently of the COBE data and any prior model for $P_k$. The galaxy-density $P_k$ by Peacock and Dodds is shown in comparison. The figure also shows the power spectra of three other popular CDM models.

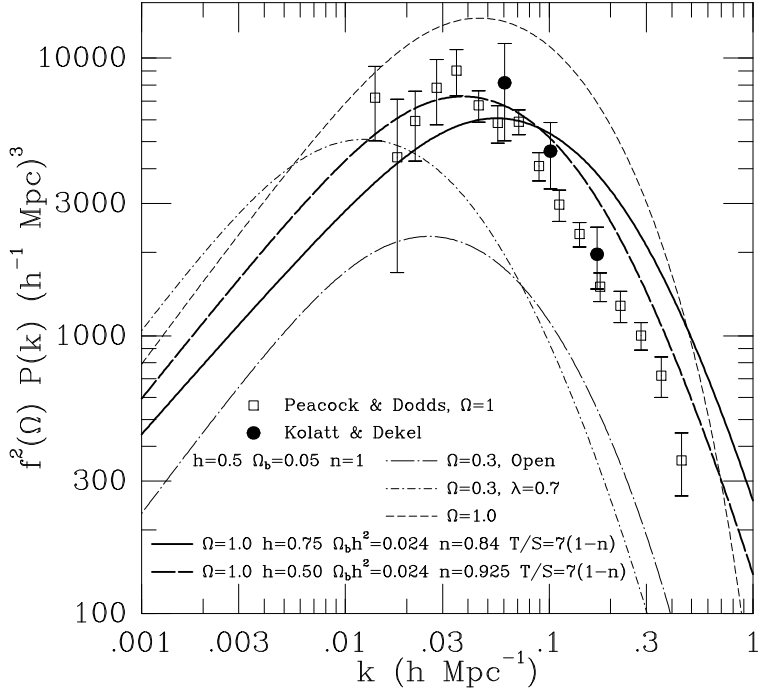

**Figure 1:** Mass power spectra in our local neighborhood. The heavy solid curve corresponds to the most probable COBE-normalized CDM model as obtained by Zaroubi *et al.* (1996) from the Mark III peculiar velocity data. The long-dashed curve corresponds to the $1\sigma$ less probable model that coincides with the best fit by White *et al.* (1995). Three popular CDM models are shown for reference. The filled dots represent the mass power spectrum as obtained by Kolatt and Dekel (1997) from the peculiar velocity data via POTENT. The open squares are the galaxy power spectrum as compiled by Peacock and Dodds (1994), for $\Omega = 1$.

### 2.2. Pre-recombination Evolution

The relation between the initial adiabatic perturbations and the temperature fluctuations at the last-scattering surface is obtained by numerically solving the Boltzmann equation (Sugiyama & Gouda 1992), using a gauge-invariant method (Bardeen 1980; Kodama & Sasaki 1984). The numerical calculations take into account only scalar perturbations. Adiabatic initial conditions are set at $T = 10^7$K, well within the radiation-dominated era. Before recombination, when the electron mean free time is shorter than the expansion time scale, the baryons and photons are treated as a single viscous fluid. Then, the Boltzmann equation is solved by expansion in multipole moments. Three species of massless neutrinos are followed by another Boltzmann equation through the whole evolution. Finally the calculation traces the evolution up to the present epoch.

During the recombination process, the detailed behavior of damping by photon diffusion (Silk 1968) is sensitive to the time evolution of the number density of free electrons.

The reionization history is solved by following Peebles (1968) and Jones & Wyse (1985). The recombination coefficient is obtained by employing the fitting formula of Pequignot, Petitjean & Boisson (1991). The difference between matter temperature and photon temperature is also taken into account. The mass fraction of Helium is assumed to be $Y = 0.23$ and the helium recombination process is treated by the Saha formula.

It is worth noting that the Boltzmann equation does not depend on the phases of the initial perturbation field. In terms of a spherical harmonic expansion this means that the solution depends only on the harmonic number $l$ (Bond & Efstathiou 1987). This makes the numerical calculations easier and faster.

### 2.3. The Radiation Power Spectrum: Comparison with CMB Data

Under the assumption that the local neighborhood is a fair sample, The acceptable range of parameters in the CDM family of modelds according to the likelihood analysis of the peculiar velocities is translated to a range in the radiation power spectrum $C_l$. It is compared to the recent CMB data under the assumption that the local neighborhood is a fair sample.

Figure 2 shows extreme cases, in terms of the height and location of the first acoustic peak of $C_l$, among the CDM models which fit the velocity data at the 90% confidence level. The four panels correspond to different CDM models. Two values of the Hubble constant $h$ are examined, low (0.5) and high (0.75). Similarly, two values of $\Omega_b h^2$ are considered, low (0.0125, Walker *et al.* 1991) and high (0.024, Tytler *et al.* 1995).

Figs. 2a and 2b show extreme cases of the tilted CDM model (TCDM), where $\Omega_m$ is fixed at unity and $n$ is free to vary, with (2a) or without (2b) tensor fluctuations. The upper two curves assume the low value of $h$ and the high value of $\Omega_b h^2$, both pushing $\Omega_b$ and therefore the peak upwards. The lower two curves assume the high $h$ and low $\Omega_b h^2$ values, thus pushing $\Omega_b$ and the peak down. The two values of $n$ define the range of 90% likelihood in $P_k$ when all the other parameters are fixed.

Fig. 2c shows the two extreme cases in the $\Lambda$CDM model, which assumes a flat universe ($\Omega_m + \Omega_\Lambda = 1$) with $n = 1$. The two values of $\Omega$ define the 90% likelihood level of $P_k$ for $h$ and $\Omega_b$ fixed as before at the values that give rise to high and low peaks.

Fig. 2d refers to the open CDM model (OCDM), where $\Lambda = 0$ and $n = 1$ are fixed. The values of $\Omega$ define the 90% likelihood range as before, but this time these are two extreme cases in the location of the acoustic peak as well as its amplitude.

We draw two main conclusions from a visual ispection of Fig. 2: (a) a wide range of CDM models that are of high likelihood in terms of the peculiar velocity data provide a good fit to the CMB data as well. (b) CDM models with too low values of $n$ and $\Omega_b$ (and too high value of $h$) can be significantly ruled out by the CMB data eventhough they are perfectly good fits to the velocity data. For example, the tilted models shown at the bottom of Fig. 2a, with $\Omega = 1$, $h = 0.75$, $\Omega_b = 0.025$, $n = 0.79 - 0.84$ and nonzero tensor component, provide an excellent fit to the peculiar velocity data, but it predicts an acoustic peak hight that is too low by at least a factor of two 2 compared to the CMB data.

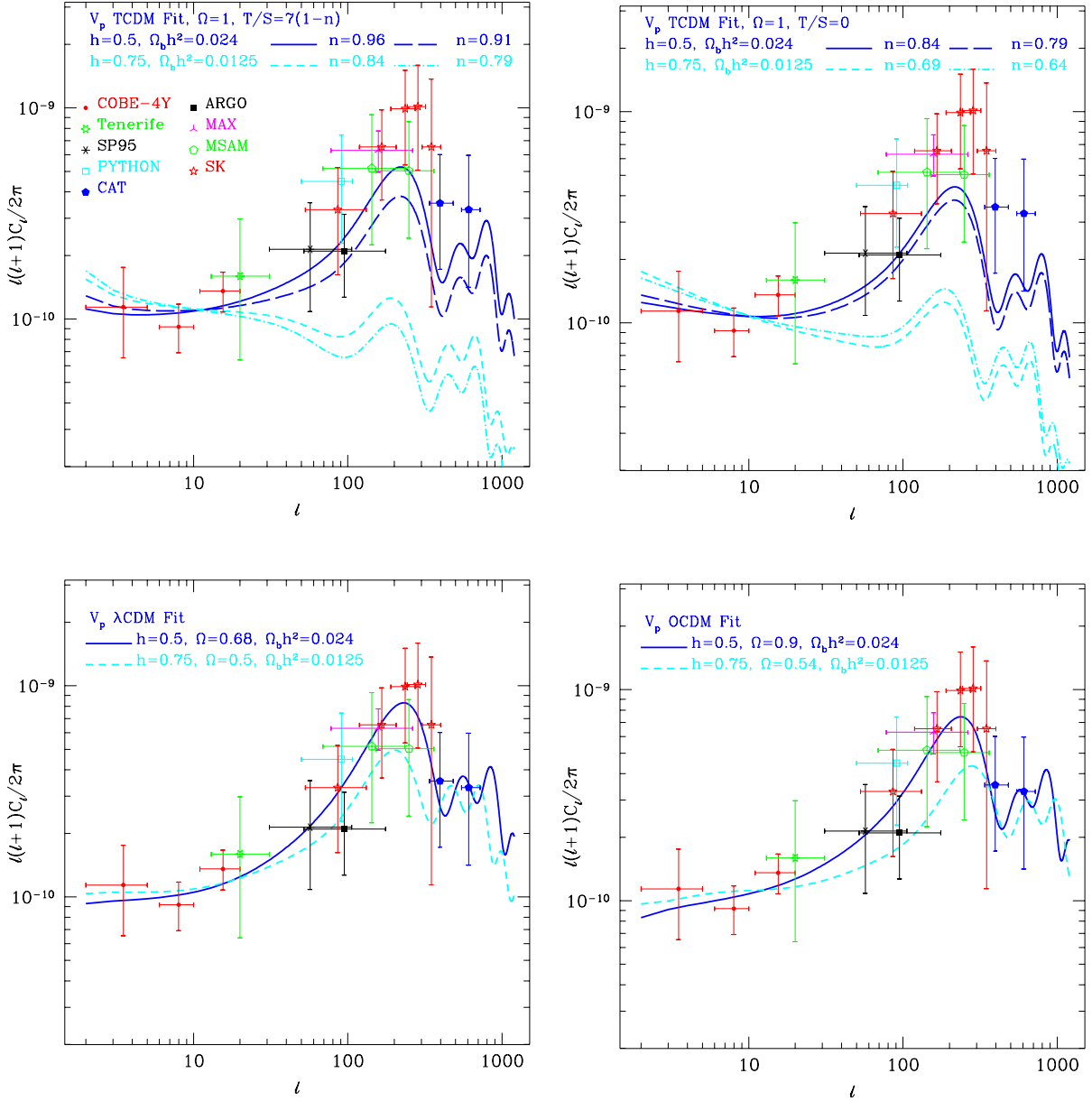

**Figure 2:** Angular power spectra of temperature fluctuations. The range of CDM models that fit the peculiar velocity data is compared to recent CMB observations (see the text). The CMB measurements are from the following experiments: 4-years COBE data (Hinshaw *et al.* 1996), 2-years COBE data (Bunn *et al.* 1995), Tenerife (Hancock etal 1996a), Python (Ruhl *et al.* . 1995), South Pole (Gundersen *et al.* 1995), ARGO (De Bernardis *et al.* 1994), MAX (Tanaka *et al.* 1996), MSAM (Cheng *et al.* 1994), Saskatoon (Scott *et al.* 1996), and CAT (Hancock *et al.* 1996b). best fit COBE-normalized CDM (heavy lines).

The fact that models which are relatively close to each other in terms of $P_k$ may be far apart in terms of $C_l$ is sometimes referred to as "cosmic confusion" (Bond *et al.* . 1996). The role of $\Omega_b$ is easy to understand: it has strong influence on the height of the $C_l$ peak, while it hardly affects $P_k$. The CMB observations prefer a high value of $\Omega_b$, while the peculiar velocities have little to add.

The role of $n$ in the cosmic confusion is also significant and is slightly more complicated. While the peculiar velocities allow values significantly lower than unity, the CMB data cannot tolerate values of $n$ below 0.9 or so. With the high value of $\Omega_b$, all the CDM models in the range $\Omega_m = 1.0 - 0.5$ and $n = 0.9 - 1.0$ respectively fit well the two data sets.

## 3. THE LOCAL NEIGHBORHOOD AS A CMB PATCH

### 3.1. Wiener Reconstruction of the Local Density Field

The peculiar velocity data allow a reconstruction of the mass density field in the local cosmological neighborhood out to $\sim 60-80 h^{-1} Mpc$. In BGD it was done via the POTENT procedure. Here we use an alternative new approach based on Wiener Filter (WF). The general application of WF to the reconstruction of large-scale structure is described in Zaroubi *et al.* (1995). The reconstruction of density field from peculiar velocity data is described in Zaroubi, Hoffman & Dekel (1997). Here we provide just a brief summary.

Let $u_i^{obs} = u_i + \epsilon_i$ be the observed radial velocity of galaxy at position $i$, where $u_i$ is the true, underlying radial velocity and $\epsilon_i$ is the error in its measurement. Assuming linear theory, the Wiener Filter provides a minimum variance estimator for the underlying (smoothed) density field, of the sort

$$\delta^{WF}(\mathbf{r}) = \langle \delta(\mathbf{r}) u_i \rangle \, \langle u_i u_j + \epsilon_i^2 \delta_{ij}^{\mathrm{k}} \rangle^{-1} u_j^{obs}, \qquad (1)$$

where $\langle \ldots \rangle$ denotes an ensemble average, $\delta$ is the smoothed density fluctuation, and $\delta_{ij}^{\mathrm{k}}$ is the Kronecker delta. The ensemble averages are calculated assuming as a prior the power spectrum of the tilted-CDM model, with $\Omega = 1$, $h = 0.75$, $n = 0.84$ and $C_2^T/C_2^S = 1.12$, which provides a good fit to the Mark III data (Zaroubi *et al.* 1996).

We note that the WF procedure provides the minimum variance solution whether the fields are Gaussian or not. In the Gaussian case, the WF solution is also the maximum probability solution, which pushes the procedure to its fullest potential.

Figure 3a shows a contour map of the reconstructed density field in the Supergalactic plane, smoothed with a Gaussian of radius $5 h^{-1} Mpc$. The main features are marked, such as the Local Group (LG) the Great Attractor (GA), Perseus-Pisces (PP), the Sculptor Void, Virgo and Fornax. Figure 3b shows an analogous map of the density field with Gaussian smoothing of $12 h^{-1} Mpc$. This smoothing corresponds better to the angular smoothing of $10'$ FWHM that we apply to the temperature fluctuations in the CMB. Figure 3c shows the corrsponding map of the potential field.

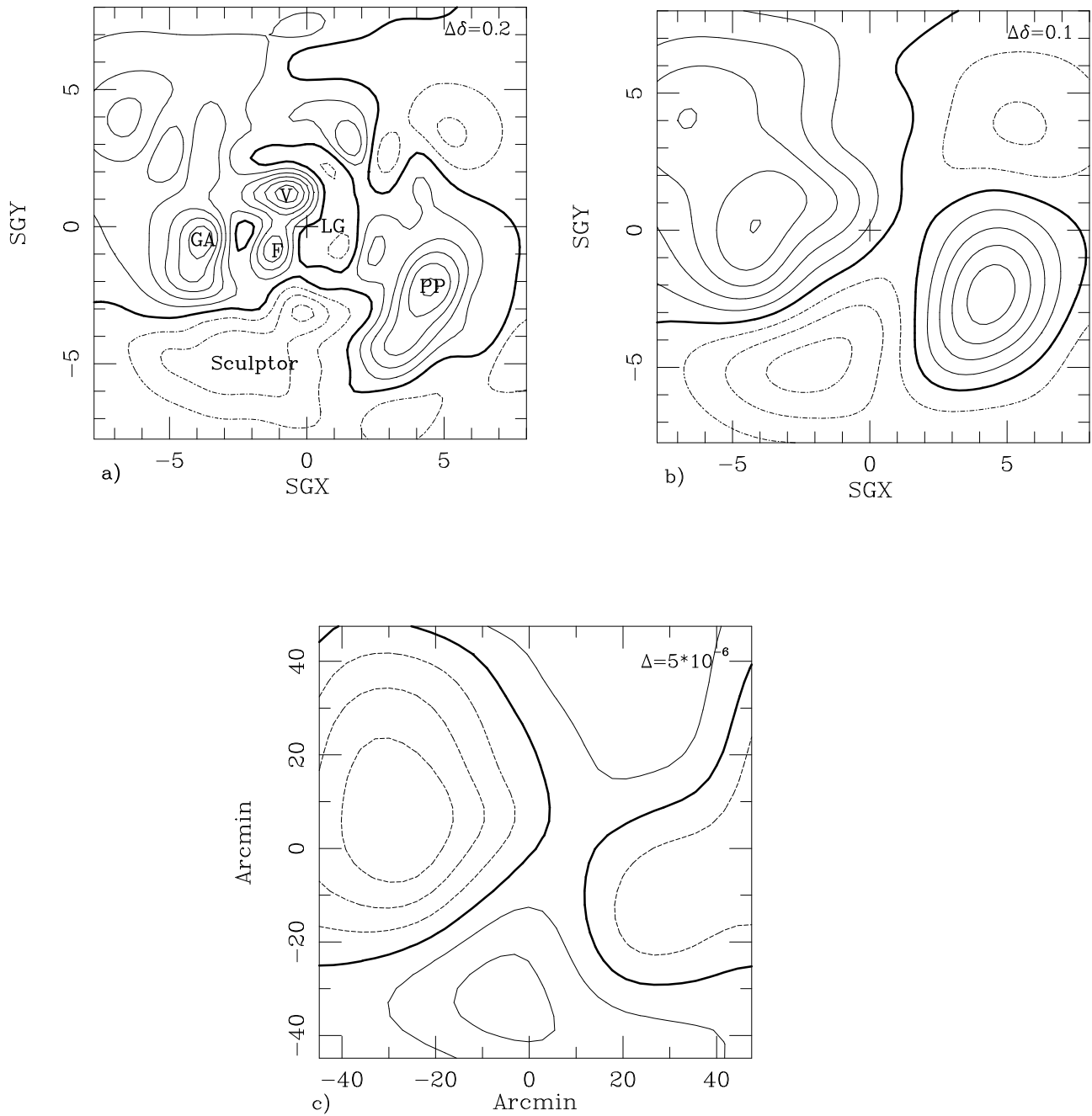

**Figure 3:** The density field ($\delta$) in the Supergalactic plane as reconstructed by Zaroubi, Hoffman and Dekel (1997) via Wiener analysis from the Mark III catalog of peculiar velocities. The Gaussian smoothing is of $5h^{-1}Mpc$ (3a) and $12h^{-1}Mpc$ (3b). Fig. 3c shows the corresponding map of gravitational potential, translated into the Sachs-Wolfe contribution to the temperature fluctuations as seen by a distant observer with $10'$ FWHM resolution ($\Omega = 1$).

### 3.2. Back in Time to the CMB

We now wish to translate the local density field into a map of temperature fluctuations as viewd by a distant observer on his last-scattering surface.

The simplest way to do this, conceptually, would have been to first use linear theory to trace the local density field back in time to early epochs well before the matter-radiation equality ($z > 10^5$), and then use this density field as initial conditions for the Boltzmann equation. In order to compute $\Delta T/T$ for this specific realization of the density field, the Boltzmann equation should have been solved separately for every harmonic number $l$ and $m$ in a spherical harmonic expansion. This computation would have been very time consuming.

We adopt instead a more economical approach, where the numerical solution of the Boltzmann equation provides a global transfer function of the sort

$$\mathcal{T}(k) \equiv \frac{\Delta T}{T}(k)/\sqrt{P(k)}, \qquad (2)$$

for a given assumed power spectrum $P(k)$ and a given cosmological model. $\Delta T/T(k)$ is the Fourier transform of the temperature fluctuation. Note that the proper phase mix is obtained by the combination of different amplitudes and signs of $\mathcal{T}(k)$. We evaluate the transfer function at $z = 800$ and 600, well after the recombination process is over (at $z \approx 1200$).

The Fourier transform of the desired temperature field is then obtained by the product of the transfer function and the Fourier transform of the actual density field as derived from the local data. The resultant 3-dimensional temperature field is smoothed with a 10' FWHM Gaussian filter, and a two-dimensional temperature map is obtained for any desired line of sight by considering the corresponding slice.

Unlike the observational angular smoothing, the smoothing along the line of sight is due to the finite thickness of the last-scattering surface. This thickness is indeed comparable to the scale defined by 10 arc-minutes. The small angular coverage of the patch (less than 1 degree) allows us to ignore the curvature of the last-scattering surface and assume a plannar slice perpendicular to the line of sight. Note that for a given density field, the reconstruction of $\Delta T/T$ does not require any assumptions about the Gaussianity of the probability distribution function.

The relation between the comoving separation on the last scattering surface and the angular separation as observed is determined by two factors. First, the geometry of space-time, as determined by the cosmological density parameter $\Omega$. In the matter-dominated era, a comoving scale $\Delta X$ corresponds to an angular separation $\Delta \theta$ that varies approximately linearly with $\Omega$ (*e.g.*, , Weinberg 1972),

$$\Delta \theta \approx 2 \arcsin(|\Delta X|\Omega/4c). \qquad (3)$$

Second, the epoch of matter-radiation equality relative to that of the recombination process. In a flat model with a large cosmological constant, $\Lambda$, the two epochs overlap. This means that during the recombination process the universe is not always matter dominated, and therefore the cosmological scale factor changes its temporal behavior along the photon path. In this case, one has to integrate over the history of the photons in order to work out the angular separation of a specific patch. This effect results in a smaller angular separation for higher values of $\Lambda$ (Hu & Sugiyama 1995a).

One test of our procedure is provided by comparing the results at $z = 800$ and $z = 600$. The CMB maps at these different times should be almost the same. The comparison between the maps indeed shows only minor differences.

Another crucial test is provided by comparing the Sachs-Wolfe (Sachs & Wolfe 1967) map as obtained by our numerical procedure with the one predicted directly from the gravitational potential (as in BGD). Figure 3c shows the Sachs-Wolfe $\Delta T/T$ map as obtained directly from the local potential field for $\Omega = 1$. In this case, the relation is simple,

$$\Delta T/T = \frac{1}{3c^2}\left(\phi_{ls} - \phi_0\right), \qquad (4)$$

where $\phi_{ls}$ is the gravitational potential at the last scattering surface and $\phi_0$ is the gravitational potential at the observer's position. We note that the phases of this map are roughly anti-correlated with the phases of the density map (Fig. 3b), as expected from a a gravitational redshift effect that is proportional to potential rather than density fluctuations. The Sachs-Wolfe map obtained by our numerical procedure when all other physical processes are shut off is found to be identical to the map shown in Fig. 3c.

We ignore any changes caused by the decay of the gravitational potential (the integrated Sachs-Wolfe effect, Sachs & Wolfe 1967), which is expected to be dominant only on scales larger than the scales of interest here (see e.g. Hu & Sugiyama 1995a).

### 3.3. CMB Maps

The contour maps of $\Delta T/T$ corresponding to our local patch as seen by a distant observer along the SGZ axis are shown in Figure 4. We have applied our procedure under three different cosmological models: (a) a flat model with $\Omega = 1$, (b) a flat model with $\Omega = 0.3$ and $\lambda = 0.7$, and (c) an open model with $\Omega = 0.3$. We adopt here a low baryon content, $\Omega_b h^2 = 0.0125$ (Walker et al. 1991).

As expected, the three maps differ substantially in their angular extent and in the amplitude of the fluctuations, but the phases are quite similar.

Note that the angular extent of the patch for the $\Omega = 1$ case is about three time larger than its extent in the open $\Omega = 0.3$ case, as expected from Eq. 3 due mainly to the difference in the background geometry. Note, however, that the ratio between the angular extent in the two flat models with and without a cosmological constant is about 0.6. Here the difference is mainly due to the late occurrence of the matter-radiation equality in a flat universe with cosmological constant.

The amplitude difference between the three panels of Figure 4 can be explained by the change in the thermal history of the universe in the case of a cosmological constant (4b), and by the change of the location of the acoustic peak toward smaller scales due to the background geometry, in the case of an open model (4c) (Hu & Sugiyama 1995a).

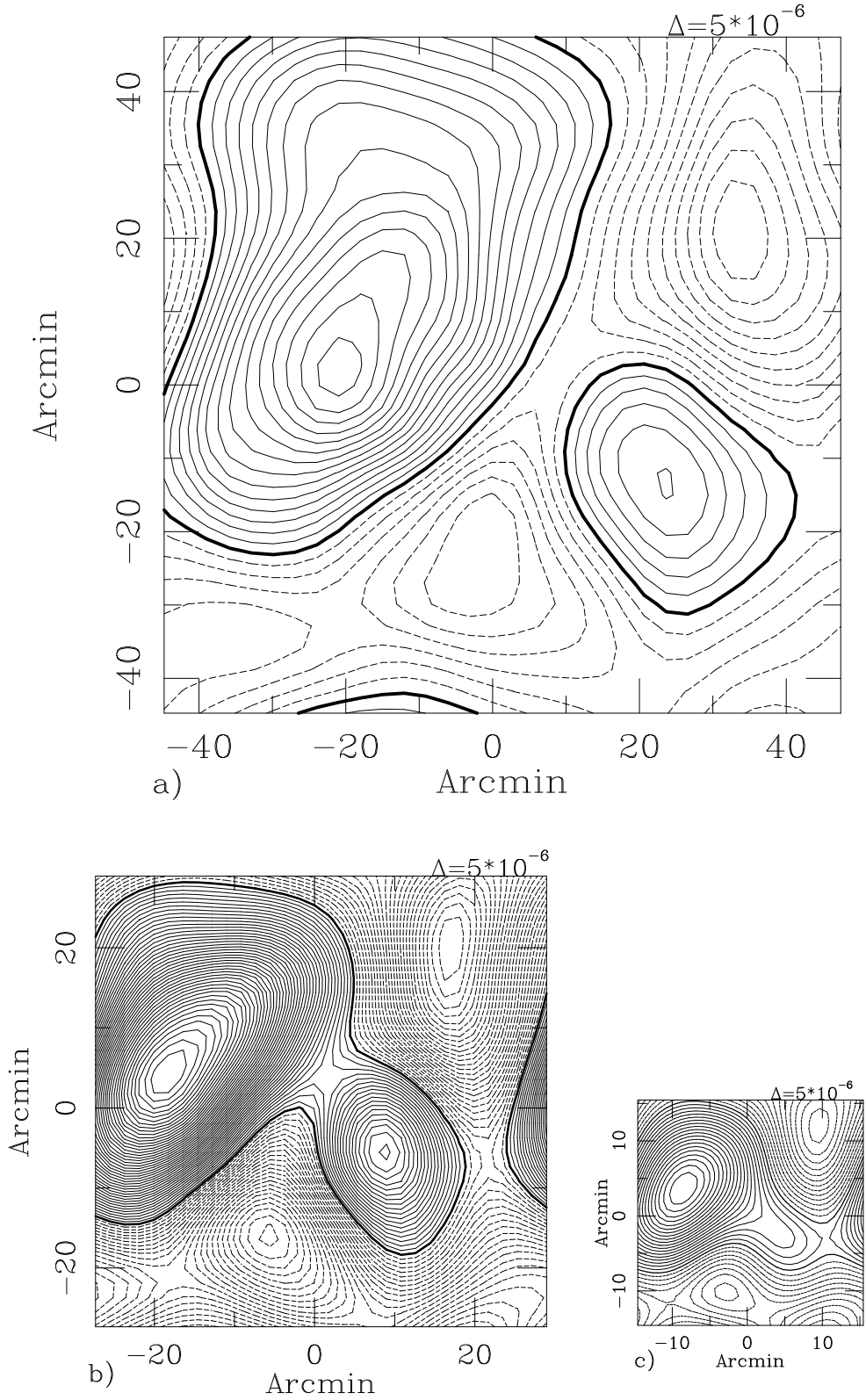

**Figure 4:** The temperature field as seen by a distant observer viewing our local neighborhood at his last-scattering surface from above the Supergalactic plane. The map are produced assuming $h = 0.5$ and $\Omega_b h^2 = 0.0125$. Contour spacing is $5 \times 10^{-6}$ in $\Delta T/T$. The maps are shown for three cosmological models: (a) $\Omega = 1$ and $\Lambda = 0$, (b) $\Omega = 0.3$ and $\lambda = 0.7$, (c) $\Omega = 0.3$ and $\Lambda = 0$.

We find a strong phase correspondence between the $\Delta T/T$ maps and the density maps (Fig. 3b). Such a correspondence is expected because the fluctuations in the CMB on these scales are caused by acoustic oscillations, which are proportional to the distribution of baryonic matter (Hu & Sugiyama 1995b; Hu, Sugiyama & Silk 1995). The latter follows the spatial distribution of dark matter on these scales. For example, the first and second peaks in the angular power spectrum (Fig. 2) are expected to be correlated and anti-correlated with the density fluctuations, respectively. To obtain the correlation seen in the maps, the correlations of the spectrum peaks should be 'convolved' with the angular extent of the smoothing scale under the assumed cosmological model.

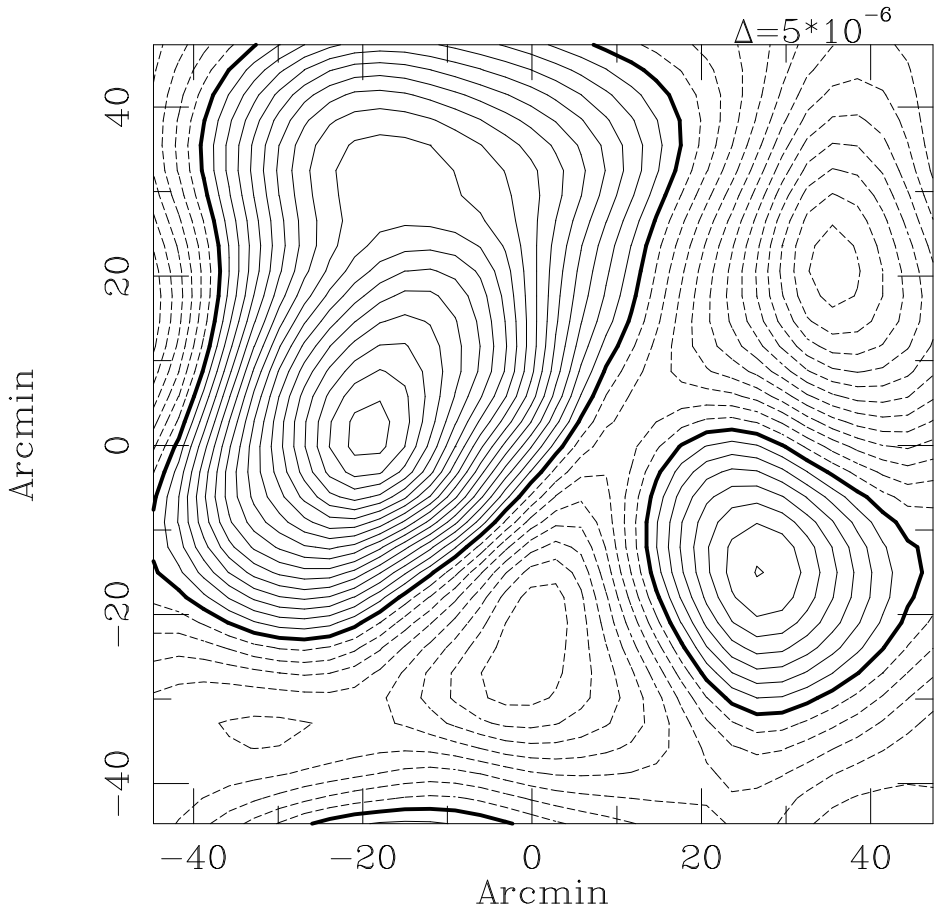

**Figure 5:** The same as Fig. 4a but with $\Omega_b h^2 = 0.025$.

In Figure 5, we show a contour map for the flat $\Omega = 1$ model but with a high value for the baryonic contribution, $\Omega_b h^2 = 0.025$ (Tytler *et al.* 1996). The fluctuation contrast becomes larger in this case compared to Fig. 4a, as expected (Sugiyama 1995). The change in $\Omega_b$ is also associated with a slight change in the phases.

Figure 6 shows three orthogonal projections of the mass density field in our local neighborhood and the associated CMB patches as viewed from the corresponding different directions ($\Omega = 1$). The variation between the projections is large. The projection onto the Supergalactic plane shows an atypically large contrast while the contrast in the other projections is more moderate. The phase correlation between the temperature and density maps is confirmed in the three projections. The coherence length in the three projections is similar, at a wavelength comparable to the patch size, $\sim 90'$.

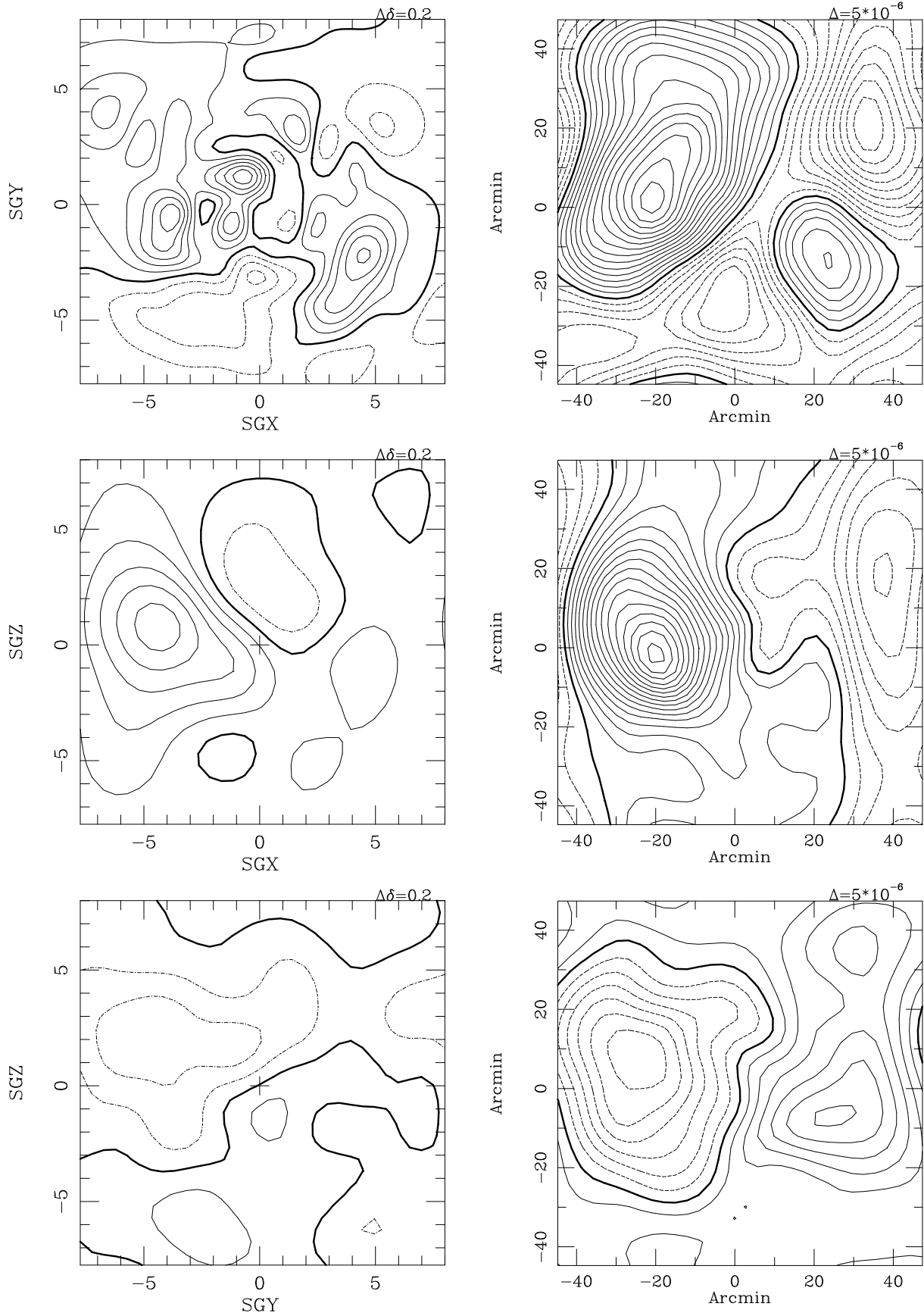

**Figure 6:** Orthogonal projections of the local density fields with $5h^{-1}Mpc$ smoothing (left) and the corresponding temperature maps with 10' smoothing (right), assuming $\Omega = 1$ and $\Lambda = 0$.

### 4. CONCLUSION

In the first part of this paper, we presented a visual comparison of theoretical and observed angular power spectra of CMB temperature fluctuations. The range of model power spectra reflects the range allowed by the Mark III data of peculiar velocities for the general family of CDM models, requiering a cosmological constant or a tilt in the initial power spectrum. This comparison assumes that the local neighborhood is a fair sample of the universe.

The comparison suggests that in order to fit both the peculiar velocity data and the CMB data the models should have a high baryon content of $\Omega_b h^2 \sim 0.025$ (as measured by Tytler *et al.* 1996), a low Hubble constant of $h \sim 0.5$, and especially a small but nonzero tilt to $n \sim 0.9$ and $0.8$ for models with and without tensor fluctuations respectively. The peculiar velocity data prefers $\Omega \geq 0.5$, and the sub-degree CMB data indicates in agreement a high value of $\Omega + \Lambda$.

The quality of the current CMB anisotropy measurments on sub-dgree scales limits us at the moment to a semi-quantitative visual comparison of the sort pursued here. The improved data expected in the near future will allow a full quantitative likelihood analysis of the two kinds of data combined.

In the second part of the paper, we presented an improved procedure for reconstructing temperature maps of sub-degree resolution from the local matter distribution, which involves mapping back in time using gravitational instability and integration of the Boltzmann equation under given cosmological models. The method was applied to the density field as reconstructed by a Wiener Filter method from the Mark III peculiar velocity data assuming three different cosmological models. A comparison of the recovered CMB maps with similar observed patches in our CMB sky can help addressing the question of whether the local neighborhood is indeed a fair sample.

The correspondence between the phase distribution of matter and radiation suggests that one can invert this procedure and, for an assumed cosmology, reconstruct the projected density map at the last-scattering surface that corresponds to the observed temperature distribution in the CMB. Projection effects (such as the integrated Sachs-Wolfe effect) can be deconvolved from the CMB fluctuations. Since this procedure is free of assumptions concerning Gaussianity, the recovered linear density field can be tested for Gaussianity at the initial conditions, with interesting theoretical implications. A statistical comparison of the density maps recovered from the CMB with the density maps in the local neighborhood on scales of $\sim 100 h^{-1} Mpc$ and beyond should enable a unique test of how well gravitational instability operates as a link between the fluctuations at $z \sim 1000$ and the present-day structure.

With a sample of local density distributions from a suitably deep survey, one could address the issue of the choice of cosmological parameters that is optimally suited to reconcile large-scale structure with the CMB anisotropies from which it originated. The spatial curvature of the universe dramatically affects the maps, and the proper volume element also has an impact via the angular size-redshift relation. One would obtain independent measures of these quantities that would complement the direct measurement of parameters from the positions and locations of the acoustic peaks in future CMB satellite experiments. This is especially important since even modest early reionization might frustrate this latter effort. Lastly, with a large complement of maps, one would be able to address the question of biasing, since one has a sample that measures the statistics of mass fluctuations (CMB) and a sample that measures the statistics of luminous mass density (redshift surveys).

**Acknowledgments:** We would like to thank W. Hu, M. Sasaki, R. Sheth and M. White for stimulating discussions. The hospitality of Racah Institue of Physics where part of this work was conducted is gratefully acknowledged (NS and SZ). This work was supported in part by the US-Israel Binational Science Foundation grants 92-00355 and 95-00330 (AD) and 94-00185 (YH), and by the Israel Science Foundation grants 469/92, 950/95 (AD) and 590/94 (YH).